# Observation of Bulk Quadrupole in Topological Heat Transport

Guoqiang Xu[1#], Xue Zhou[2#], Jing Wu[3, 4], and Cheng-Wei Qiu[1*]

[1]*Department of Electrical and Computer Engineering, National University of Singapore, Kent Ridge 117583, Republic of Singapore*

[2]*School of Computer Science and Information Engineering, Chongqing Technology and Business University, Chongqing, 400067, China*

[3]*Institute of Materials Research and Engineering, Agency for Science, Technology and Research, Singapore, Singapore*

[4]*Department of Materials Science and Engineering, National University of Singapore, Singapore, Singapore*

[#]These authors contributed equally.

[*]Correspondence to: <u>chengwei.qiu@nus.edu.sg</u>

**Abstract**

Quantized bulk quadrupole moment has unveiled a nontrivial boundary state, exhibiting lower-dimensional topological edge states and simultaneously hosting the in-gap corner modes of zero dimension. All state-of-the-art strategies for topological thermal metamaterials have so far failed to observe such higher-order hierarchical features, since the absence of quantized bulk quadrupole moments in thermal diffusion fundamentally forbids the possible expansions of band topology, unlike its photonic counterpart. Here, we report a recipe of creating quantized bulk quadrupole moments in diffusion, and observe the quadrupole topological phases in non-Hermitian thermal systems. The experiments demonstrate that both the real- and imaginary-valued bands showcase the hierarchical hallmarks of bulk, gapped edge, and in-gap corner states, in stark contrast to the higher-order states only observed on real-valued bands in classic wave fields. Our findings open up unique possibilities for diffusive manipulations and establish an unexplored playground for multipolar topological physics.

Topological states of matter have found explosive developments across various classic wave fields [1-5]. In an adiabatic system, hermiticity lies at the foundation of these emerging topological properties [6, 7], as it ensures the real-valued eigenvalues and orthogonal eigenstates. When considering open systems, additional interactions with ambient raise the non-hermiticities. Though these dissipations fail the fundamental bulk-boundary correspondence [8, 9] defined in Hermitian system, a plethora of exotic properties are empowered, such as parity-time symmetry [10-12], skin effects [13, 14], as well as Weyl exceptional rings in cold atomic gas [15], photonics [16], and semimetal [17]. The newly predicted higher-order topological insulators (HOTI) have further paved an avenue toward studying hierarchical features in both Hermitian [18-23] and non-Hermitian [24-27] systems. Featuring a quantized bulk quadrupole moment [18], the Benalcazar-Bernevig-Hughes (BBH) model holds the key for realizing a minimal quadrupole topological insulator (QTI) possessing positive and negative couplings [18-20]. Moreover, a modified non-Hermitian BBH model indicates that both the on-site non-hermiticities [24] and the hermiticities [25] can derive the quadrupole topological phases and modulate the higher-order transitions in real-valued bands [24-27].

It is recently found that dissipative diffusion is fundamentally governed by skew-Hermitian physics and characterized by a purely imaginary Hamiltonian [28, 29]. It thus enables the counter-intuitive topological features in heat transport, such as non-Hermitian topological insulating phases [30] and Weyl exceptional rings [31]. On the other hand, even the state-of-the-art methods [28-31] fail to create non-Hermitian thermal quadrupole topological phases, due to the absent bulk quadrupole moment and undefined negative couplings in heat transfer. Therefore, to date, non-Hermitian BBH model seems not applicable to heat transport, and quadrupole topological phases in thermal diffusion are still elusive at large.

Here, we reveal the existence of quadrupole moment and non-Hermitian quadrupole topological phases in heat transport for the very first time. It is essentially realized by judicious configurations of controllable thermal couplings between neighboring sites. In contrast to the fact that the higher-order features can only be observed on real-valued bands in classic wave fields, we capture these states on both real- and imaginary-valued bands. We then experimentally demonstrate these thermal quadrupole topological phases, and observe significant temperature localizations at the bulk, edge, and corner of the fabricated samples. Our work sheds new light upon establishing quantized bulk quadrupole moments in thermal systems and unlocking rich topological phase transitions in various diffusions [32-35].

We first consider a convective thermal system with multiple discrete sites as illustrated in Figs. 1**a** and **b**. Each site indicates a finite-volume of heat transfer process, and the grid lines between neighboring sites correspond to their thermal couplings. We adopt tunable advections on each site to provide the necessary modulation and create effective oscillations (i.e., resonances) in the thermal system, thus further forming an effective unit-structure consisting of four neighboring sites (Fig. 1**b**). Such unit-structures can be periodically configured to establish an effective 2D square-lattice with 16 sites (Fig. 1**a**) in heat transfer. The general heat transfer process of each site can be expressed as

$$\frac{\partial T_{ij}}{\partial t} = \frac{\kappa}{\rho c}\nabla^2 T \pm \omega_{I/II} R(\theta)\nabla T + \sum \beta h \Delta T. \tag{1}$$

In Eq. (1), $\rho$, $c$, and $\kappa$ respectively denote the density, specific heat, and thermal conductivity of the site. Each site is depicted by its position ($i$ and $j$), and $T_{ij}$ denotes the corresponding temperature (Supplementary Figure 1**a**). $\Omega_{I/II}$ represents the magnitude of the angular velocities of the convection imposed on each site, and $R$ and $\theta$ respectively denote the radial and azimuth components in the $x$-$y$ plane. $h$ indicates the heat transfer coefficient of the selected site, and $a_{x/y}$ presents the widths for heat transport between the centers of neighboring sites (Fig. 1**b**). Here we make $a_x = a_y = a$ in the current system to ensure an effective square unit-structure. $\beta$ is the ratio between the intercell and intracell thermal coupling strengths, and its value is 1 when the heat exchange areas of the intracell and intercell components are same (**Methods**). Taking into account thermal couplings in such a 2D network, two components along the $x$ and $y$ directions can be decoupled from the imposed advections on each site, i.e., $\Omega_{I/II} \cdot \cos\theta$ and $\Omega_{I/II} \cdot \sin\theta$. In that case, a diffusive analog to the quantized bulk quadrupole moment could emerge and exhibit characteristic quadrupole fields in temperature distributions when modulating the advection and thermal coupling in the unit-structure (Supplementary Note 3). That further enables the effective dipole moments on 1D edges and the uncompensated charges on 0D corners.

Since the intrinsic thermal coupling is governed by skew-Hermitian physics [27-30], the imposed advections act as the real hermiticities, which are equivalent to the roles of gain and loss in photonics. Note that the neighboring sites are coupled via the heat exchanges induced by intracells and intercells. The titled connections (Supplementary Figure 2) between two adjacent sites result in different orientations of isotherms and coupling degrees under the same advections (Supplementary Figure 3). Such an implementation enables the over-coupling and under-coupling, with respect to the reference coupling strength in un-tilted configurations (Supplementary Note 3). The wave-like solution on each site reveals the oscillatory propagation of the temperature fields.

The thermal couplings between neighboring sites originate from both intracell and intercell heat exchanges. The intercell couplings should follow the Bloch theorem, where $k_{t,x}$ and $k_{t,y}$ are the directional components of the effective Bloch wave number $k_t$ (Supplementary Note 1). The complex angular frequency and eigenvalues imply that both the advections and corresponding couplings result in the complex bands. The imaginary angular frequency originates from the intrinsic conduction and the thermal couplings, while the real angular frequency represents the effective momentum induced by the imposed advections towards different azimuths. These two parts simultaneously determine the amplitudes and the movements of the dynamic temperature field, thus retaining the possibilities of exciting significant hierarchical states with two distinct recipes, i.e., modulating the Hermitian advection and the non-Hermitian coupling. We then fabricate a square-lattice with 16 sites and immerse it into water (**Methods**). All fabricated sites are hollow in order to make water pass through and connect with tilted channels possessing tailored thermal coupling strengths. These sites are of the same size and act as advective balls to provide the needed advections.

We first focus on the quadrupole topological phases enabled by the Hermitian advection. The real phase diagrams of the first Brillouin zones under specific advections are presented in Fig. 1**c**. When the imposed advective magnitudes are same and the directions are opposite ($\Omega_I = -\Omega_{II}$), all the real-valued bands degenerate indicating the gapless structure and the transition between nontrivial and trivial thermal states. When we slightly increase $\Omega_{II}$ to $-1.385\Omega_I$ to maintain an advective difference, two gaps emerge between four bands, and the second and third bands are still degenerate. Further increasing $\Omega_{II}$ to $-3.154\Omega_I$, all degeneracies lift and a complete open band emerges. Such a transition from a gapless phase to a gapped phase via solely modulating the Hermitian advection represents a class of topological quadrupole phases, embracing the in-gap 0D and gapped 1D topological modes. We then calculate the dispersion relations to further validate the existence of these quadrupole topological phases. The robust in-gap corner state (red dots) and gapped edge state (blue dots) are respectively presented in Figs. 1**d** and **e**, revealing these higher-order states with the advective configurations of $\Omega_{II} = -1.385\Omega_I$ and $\Omega_{II} = -2.077\Omega_I$. The completely gapped bands illustrated in Fig. 1**f** with $\Omega_{II} = -3.154\Omega_I$ present a thermal analog of a trivial insulator.

We then fabricate a thermal system consisting of 12 sites (9 square lattices) along the *x* and *y* directions (Fig. 2**a**) to manifest these nontrivial states. In order to ensure the topological transitions solely via the Hermitian advection (Supplementary Equation 9), the same structures are adopted in all coupling channels to retain the same intercell and intracell thermal coupling strength ($\beta$= 1). One of the imposed advections on a pair of diagonal sites in one unit-structure (Fig. 2**a**) is adopted as $\Omega_I = 1.3Q_c$ based on the calculated dispersion in Fig. 1**d**, while we sweep advection $\Omega_{II}$ on the other pair of diagonal sites to search for corresponding real angular frequency (Supplementary Equation 9). Due to the effective quantized quadrupole moment enabled by the above advective arrangements (Supplementary Notes 2 ~ 4), the eigenfrequency spectrum indicates that significant hierarchical features discretely distribute along the real-valued band and localize on one gapless imaginary-valued band (Fig. 2**b**). When Re$f$ respectively approaches 0 and $4.81Q_c$, the trivial bulk states showcase the gaps between these two branches in the real-valued band. We choose three sites respectively at the center, edge, and corner of the sample (marked as a square in Fig. 2**a**), and capture their responses under changing Re$f$ as plotted in Fig. 2**c**. Here, we take the absolute values of the normalized temperatures $\left|\frac{(T^* - \bar{T}_{mea})}{\Delta T_{mea}}\right|$ to evaluate field intensities, where $T^*$, $\bar{T}_{mea}$ and $\Delta T_{mea}$ respectively denote the target temperature at specific measured points, the average temperature of the system, and the difference between the highest temperature and $\bar{T}_{mea}$ at the measured moments. Two peaks of the field intensities are observed at corresponding Re$f$ to the bulk branches as predicted in Fig. 2**b**. Similar to the responses in the bulk, the gapped edge states also exhibit two peaks as the gradient blue area in Fig. 2**c**. The four in-gap corner states emerge. In that case, only one peak is found on the field intensity distribution. To further experimentally demonstrate these quadrupole topological phases, we measure the temperature distributions by modulating the advections as shown in Figs. 2**d** ~ **f**. When adjusting $\Omega_{II}$ to the corresponding peaks of edge and corner states, the field intensities would be larger at the corner and edge as illustrated in Figs. 2**g** and **h**. By setting the advections $\Omega_I = 1.3Q_c$ and $\Omega_{II} = -3.154\Omega_I$ (bulk state), the field intensities in the central domain are much higher than the ones at the corners and edges (Fig. 2**i**). These theoretical and experimental findings reveal the quadrupole topological phases in real-valued bands solely induced by Hermitian advection in a thermal system.

Importantly, we demonstrate that such quadrupole topological phases can also be enabled by the intrinsic non-hermiticities and captured along the imaginary-valued bands (Supplementary Equation 10). Note that, such states in these imaginary-valued bands can be theoretically observed either in a skew-Hermitian thermal system without advections ($\Omega_I = \Omega_{II} = 0$) or a non-Hermitian heat transfer with advections possessing the same magnitudes and direction ($\Omega_I = \Omega_{II} \neq 0$). Here, we focus on the non-Hermitian strategy and further demonstrate the quadrupole topological phases as illustrated in Fig. 3**a** ($\Omega_I = \Omega_{II} = 0.025Q_c$). In that case, the real-valued band is gapless and can be adopted to distinguish the states along the gapped imaginary-valued band (Supplementary Note 4). The intracell and intercell thermal coupling strengths should be also different at this stage, since the same coupling strengths would otherwise close the imaginary-valued bands and indicate a trivial bulk state [24, 25] instead. The coupling strengths can be manipulated by the heat exchanges within the intercell and intracell channels. For simplification, we keep the same intercell coupling channels as the case shown in Fig. 1. We further modify the structure by enlarging the intracell coupling channels (the right lower inset of Fig. 3**a**) and inserting internal fins (Fig. 3**b**). Such implementations lead to stronger intracell thermal couplings under the same energy inputs and enable the modulations of $\beta$ ranging from 0 to 1. The imaginary phase diagrams of the first Brillouin zones of one modified square-lattice are presented in Supplementary Figure 5**a**. Similar to the modulations with Hermitian advection (Fig. 1**c**), all the imaginary-valued bands degenerate with the same intracell and intercell coupling strengths ($\beta = 1$). Two gaps (one between the first and second bands, and the other between third and fourth bands) are observed when modulating $\beta$, thus revealing the 1D edge and 0D corner states in the imaginary-valued bands. The dispersion relations further validate the existences of in-gap corner (Fig. 3**c**), gapped edge (Fig. 3**d**), and trivial bulk (Fig. 3**e**) states along the imaginary-valued bands at tailored $\beta$.

In what follows, we construct the thermal system with 9 modified square-lattices as illustrated in Fig. 4**a** and Supplementary Figure 5**b**. When $\beta$ respectively approaches near-zero and 1 in the experiments, two branches are localized along Im$f$ and imply the trivial bulk states (Fig. 4**b**). When selecting $\beta$ in the range of 0 to 1, two gapped edge and one in-gap corner states are also expected. The field intensities on imaginary-valued bands (Fig. 4**c**) further verify the above hypothesis with two peaks on the central/edge and one peak on the corner of the measured sites (Supplementary Figure 5**b**). Such features overlap well with the experimental temperature field distributions (Figs. 4**d** ~ **f**) and the measured intensity profiles (Figs. 4**g** ~ **i**) under corresponding $\beta$. These results (Figs. 2 and 4) demonstrate the proof-of-concept quadrupole topological phases in non-Hermitian thermal systems via controlling either the imposed advections (hermiticity) or the thermal coupling strengths (non-hermiticity).

We report the creation of an effective quadrupole moment in heat transport and observe the non-Hermitian thermal quadrupole topological phases. Our results highlight the fundamental properties of these higher-order diffusive quadrupoles that drastically deviate from the wisdom about HOTIs in classic wave fields. The complex eigenvalues enable the phase transitions on both the real- and imaginary-valued bands. By modulating either the Hermitian advection or the non-Hermitian thermal coupling, the experimental demonstrations exhibit significant hierarchies of topological states in heat transport. Quadrupole topological phases in diffusive domains may reveal exotic physics on complex bands and empower the topological diffusion in fractal systems [36] and moiré lattices [37, 38]. These diffusive bulk, edge and corner states as discovered in this work may further shed new lights on the control of mass concentration in biomedicine and catalysis as well as the charge diffusion in semiconductors, and many other diffusive fields at large.

**Acknowledgements**

C.-W.Q. acknowledged the financial support by Ministry of Education, Republic of Singapore (Grant No.: A-8000107-01-00). X. Z. acknowledged the financial support of Chongqing Natural Science Foundation (Grant No. cstc2021jcyj-msxmX0627) and the Science and Technology Research Program of Chongqing Municipal Education Commission (Grant No. KJQN202000829). J. W. acknowledges the SERC Central Research Fund and Advanced Manufacturing and Engineering Young Individual Research Grant (AME YIRG Grant No.: A2084c170).

**Author contributions**

G.X. and C.-W.Q. conceived the idea. G.X., X.Z., and C.-W.Q. proposed the methodology. G.X. performed the theoretical derivation, and G.X., and X.Z. implemented the experimental investigations. G.X., X.Z., J.W., and C.-W.Q. made the visualizations. G.X., J.W., and C.-W.Q. performed the theoretical analysis and wrote the manuscript. C.-W.Q. supervised the work. All authors contributed to the discussion and finalization of the manuscript.

**Competing interests**

The authors declare no competing interests.

**Data Availability Statement**

All data that support the findings of this study are available from the corresponding author upon reasonable request.

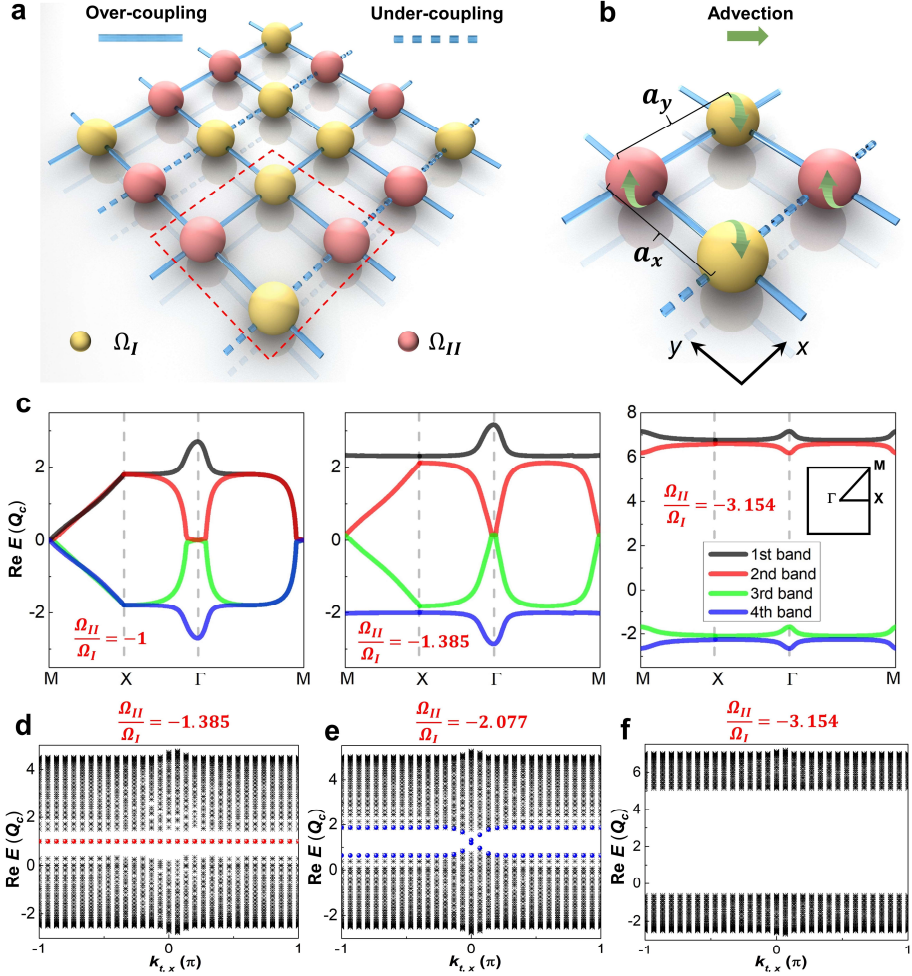

**Fig .1.** Quadrupole topological phases in heat transport induced by Hermitian advection and phase diagrams. **a** presents a square-lattice consisting of 16 sites in a grid thermal system. The red border indicates a four-site unit-structure. **b.** Schematic unit-structure with four sites. The light-yellow and light-red colors indicate the counter-advections imposed on corresponding sites, and their magnitudes are $\Omega_I$ and $\Omega_{II}$. The green arrows present the advective directions. Each advection can be decoupled as two advective components respectively in the *x-z* and *y-z* planes. The solid and dashed lines respectively present the over-coupling and under-coupling channels between neighboring sites, whose intercell and intracell coupling strengths are same. **c** denotes the real-valued phase diagrams of the first Brillouin zones under specific advections. **d ~ f** plot the real spectra of the dispersion induced by Hermitian advection respectively at the in-gap corner, gapped edge, and trivial bulk states.

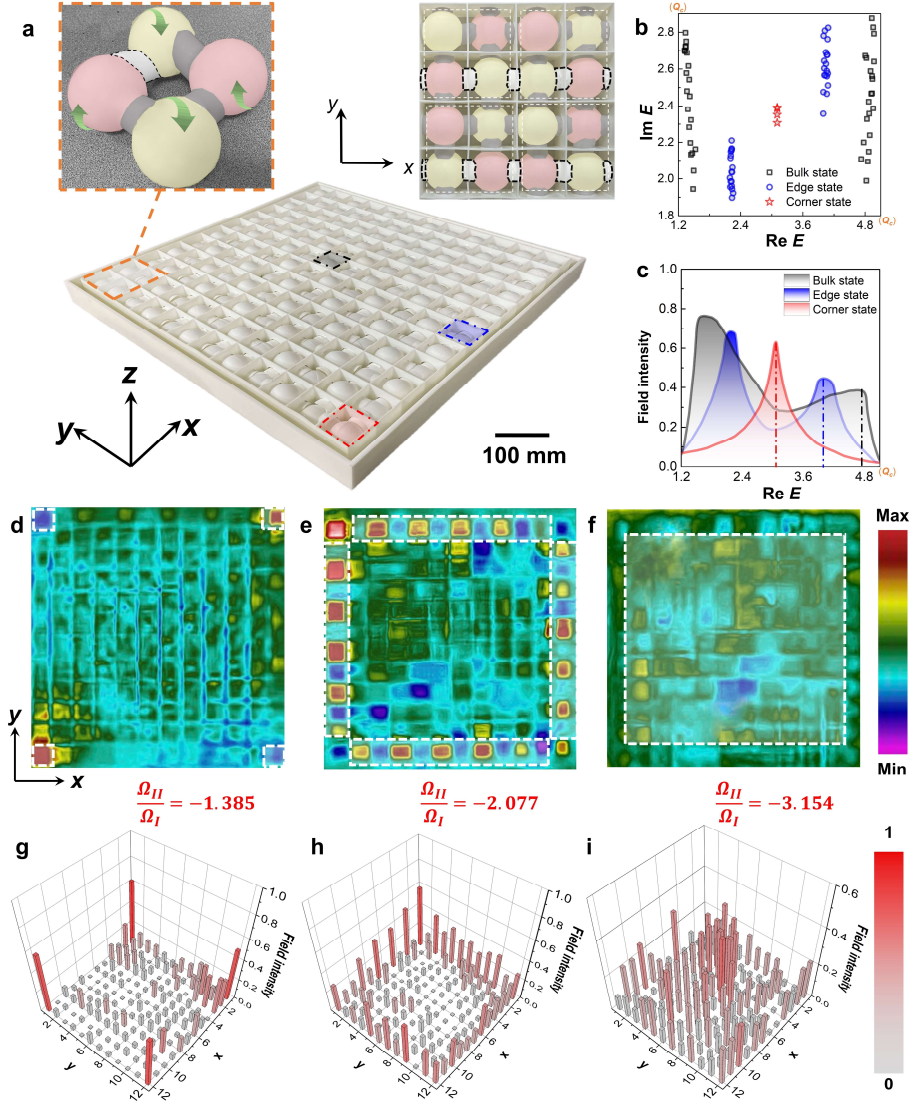

**Fig. 2.** Quadrupole topological phases in heat transport solely induced by Hermitian advection. **a** Photo of a fabricated sample with 9 square-lattices (144 sites) possessing the same intercell and intracell coupling channels. The left and right upper insets present the connection of a fabricated unit-structure and one square lattice with 16 sites. The grey-shadowed and black-dashed areas respectively indicate the over-coupling and under-coupling channels. **b** Calculated spectrums of the thermal quadrupole topological phases in (**a**). The hierarchical topological states discretely distribute along the real-valued band related to the advective differences. **c** Measured temperature field intensities at corresponding boundaries. The measured regions are marked by colored borders in (**a**). **d ~ f** Captured temperature distributions at the peaks of the corner, edge, and bulk spectra as indicated by the red, blue, and black dashed lines in (**c**). **g ~ i** Temperature field intensity profiles corresponding to the cases in (**d**) ~ (**f**).

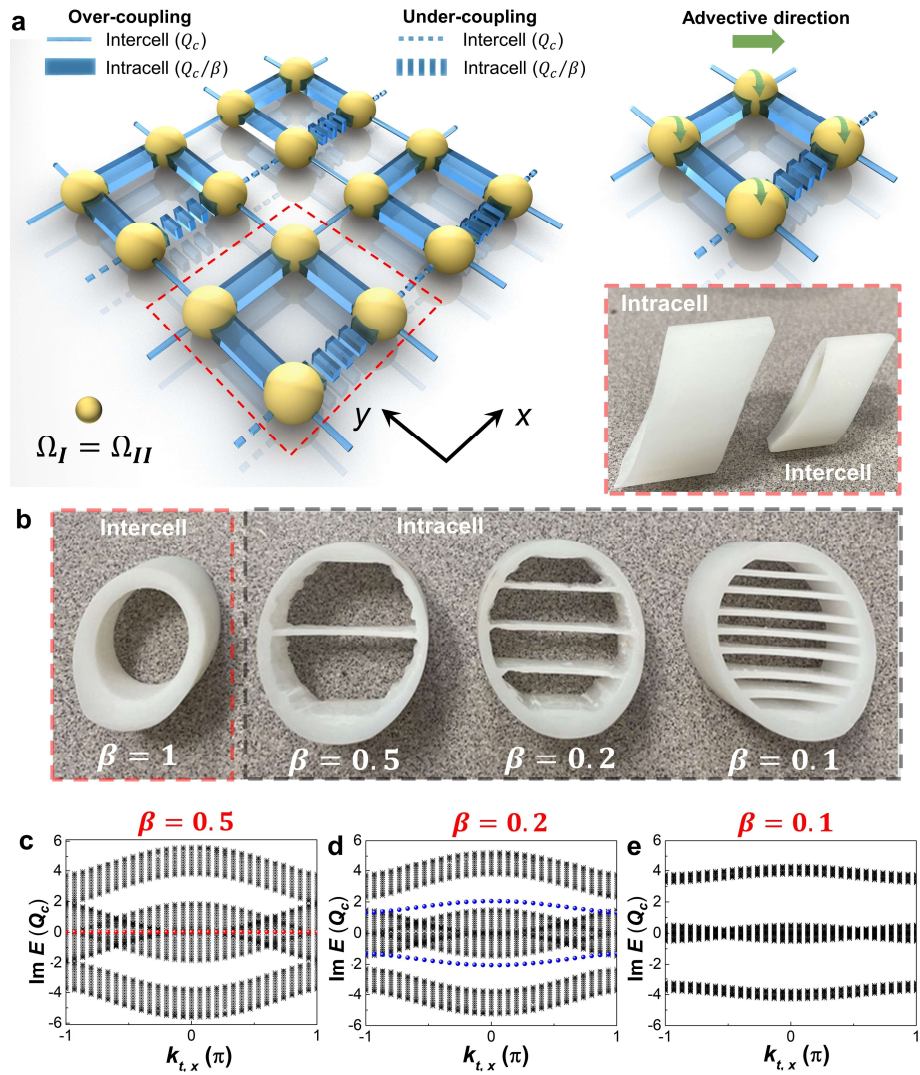

**Fig. 3.** Quadrupole topological phases in heat transport induced by non-Hermitian couplings and phase diagrams. **a** illustrates the square-lattice possessing 16 sites with different coupling strengths and the four-site unit structure. The directions and magnitudes of the imposed advections (green arrows) on each site are same to hold the non-Hermitian properties. The intercell and intracell channels are fabricated to different structures to enable the different thermal coupling strengths. **b** presents the inner structures of these thermal coupling channels with different coupling strength ratios. **c ~ e** indicate the imaginary spectra of the dispersion induced by non-Hermitian couplings respectively at the in-gap corner, gapped edge, and trivial bulk states.

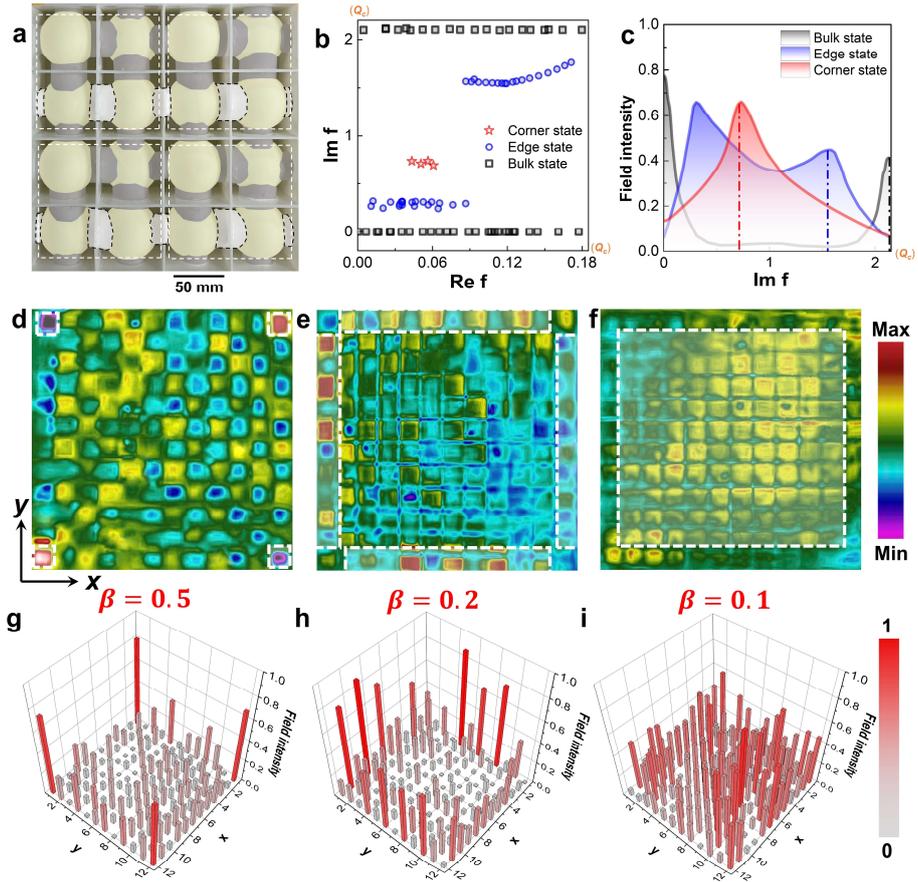

**Fig. 4.** Observation of the quadrupole topological phases in heat transport solely induced by non-Hermitian couplings. **a** Photo of a fabricated square-lattice (16 sites) with different intercell and intracell coupling channels. The grey-shadowed and black-dashed areas respectively indicate the over-coupling and under-coupling channels. Supplementary Figure 5b presents the entire sample. **b** Calculated spectrums of the thermal quadrupole topological phases induced by non-Hermitian couplings. The hierarchical topological states discretely distribute along the imaginary-valued band related to the thermal coupling strength ratios $β$. **c** Measured temperature field intensities at corresponding boundaries. The measured regions are marked by colored borders in Supplementary Figure 5b. **d ~ f** Captured temperature distributions at the peaks of the corner, edge, and bulk spectra as indicated by the red, blue, and black dashed lines in (**c**). **g ~ i** Temperature field intensity profiles corresponding to the cases in (**d**) ~ (**f**).